

\input jnl.tex

\catcode`@=11
\newcount\r@fcount \r@fcount=0
\newcount\r@fcurr
\immediate\newwrite\reffile
\newif\ifr@ffile\r@ffilefalse
\def\w@rnwrite#1{\ifr@ffile\immediate\write\reffile{#1}\fi\message{#1}}

\def\writer@f#1>>{}
\def\referencefile{
  \r@ffiletrue\immediate\openout\reffile=\jobname.ref%
  \def\writer@f##1>>{\ifr@ffile\immediate\write\reffile%
    {\noexpand\refis{##1} = \csname r@fnum##1\endcsname = %
     \expandafter\expandafter\expandafter\strip@t\expandafter%
     \meaning\csname r@ftext\csname r@fnum##1\endcsname\endcsname}\fi}%
  \def\strip@t##1>>{}}

\def\citeall#1{\xdef#1##1{#1{\noexpand\cite{##1}}}}
\def\cite#1{\each@rg\citer@nge{#1}}     

\def\each@rg#1#2{{\let\thecsname=#1\expandafter\first@rg#2,\end,}}
\def\first@rg#1,{\thecsname{#1}\apply@rg}       
\def\apply@rg#1,{\ifx\end#1\let\next=\relax
\else,\thecsname{#1}\let\next=\apply@rg\fi\next}

\def\citer@nge#1{\citedor@nge#1-\end-}  
\def\citer@ngeat#1\end-{#1}
\def\citedor@nge#1-#2-{\ifx\end#2\r@featspace#1 
  \else\citel@@p{#1}{#2}\citer@ngeat\fi}        
\def\citel@@p#1#2{\ifnum#1>#2{\errmessage{Reference range #1-#2\space is bad.}%
    \errhelp{If you cite a series of references by the notation M-N, then M and
    N must be integers, and N must be greater than or equal to M.}}\else%
 {\count0=#1\count1=#2\advance\count1
by1\relax\expandafter\r@fcite\the\count0,%

  \loop\advance\count0 by1\relax
    \ifnum\count0<\count1,\expandafter\r@fcite\the\count0,%
  \repeat}\fi}

\def\r@featspace#1#2 {\r@fcite#1#2,}    
\def\r@fcite#1,{\ifuncit@d{#1}
    \newr@f{#1}%
    \expandafter\gdef\csname r@ftext\number\r@fcount\endcsname%
                     {\message{Reference #1 to be supplied.}%
                      \writer@f#1>>#1 to be supplied.\par}%
 \fi%
 \csname r@fnum#1\endcsname}
\def\ifuncit@d#1{\expandafter\ifx\csname r@fnum#1\endcsname\relax}%
\def\newr@f#1{\global\advance\r@fcount by1%
    \expandafter\xdef\csname r@fnum#1\endcsname{\number\r@fcount}}

\let\r@fis=\refis                       
\def\refis#1#2#3\par{\ifuncit@d{#1}
   \newr@f{#1}%
   \w@rnwrite{Reference #1=\number\r@fcount\space is not cited up to now.}\fi%
  \expandafter\gdef\csname r@ftext\csname r@fnum#1\endcsname\endcsname%
  {\writer@f#1>>#2#3\par}}

\def\ignoreuncited{
   \def\refis##1##2##3\par{\ifuncit@d{##1}%
     \else\expandafter\gdef\csname r@ftext\csname
r@fnum##1\endcsname\endcsname%

     {\writer@f##1>>##2##3\par}\fi}}

\def\r@ferr{\endreferences\errmessage{I was expecting to see
\noexpand\endreferences before now;  I have inserted it here.}}
\let\r@ferences=\references
\def\references{\r@ferences\def\endmode{\r@ferr\par\endgroup}}

\let\endr@ferences=\endreferences
\def\endreferences{\r@fcurr=0
  {\loop\ifnum\r@fcurr<\r@fcount
    \advance\r@fcurr by 1\relax\expandafter\r@fis\expandafter{\number\r@fcurr}%
    \csname r@ftext\number\r@fcurr\endcsname%
  \repeat}\gdef\r@ferr{}\endr@ferences}


\let\r@fend=\endpaper\gdef\endpaper{\ifr@ffile
\immediate\write16{Cross References written on []\jobname.REF.}\fi\r@fend}

\catcode`@=12

\citeall\refto          
\citeall\ref            %
\citeall\Ref            %

\def\pd{\partial}
\def\slash#1{\raise.15ex\hbox{$/$}\kern-.57em\hbox{$#1$}}
\def\bra#1{\langle#1\vert}      
\def\ket#1{\vert#1\rangle}      
\def\ipr#1#2{\langle#1\vert#2\rangle} 
\def\me#1#2#3{{\langle#1\vert#2\vert#3\rangle}} 

\def\de{{\delta}}
\def\cD{{\cal D}}
\def\cF{{\cal F}}
\def\cL{{\cal L}}
\def\cH{{\cal H}}
\def\cN{{\cal N}}

\def\cZ{{\cal Z}}
\def\slD{\raise.15ex\hbox{$/$}\kern-.57em\hbox{$D$}}
\def\slp{{\raise.15ex\hbox{$/$}\kern-.57em\hbox{$\partial$}}}
\def\lnA{\raise.15ex\hbox{$/$}\kern-.57em\hbox{$A$}}
\def\lnB{\raise.15ex\hbox{$/$}\kern-.57em\hbox{$B$}}

\rightline{June 18, 1992}

\title Ground state wave functionals for $1+1$-dimensional fermion field
theories
\author Eduardo Fradkin and Enrique Moreno
\affil Department of Physics
       University of Illinois at Urbana-Champaign
       1110 W.Green St., Urbana, IL 61801, USA
\centerline{and}
\author Fidel A.Schaposnik
\affil Departamento de F{\'\i}sica
       Universidad Nacional de La Plata
       C.C.67, 1900 La Plata, Argentina
       and
       Comision de Investigaciones Cientificas, Buenos Aires, Argentina.

\abstract{
We use path-\-integral methods to derive the
ground state wave functions of a number of two-\-dimensional
fermion field theories and related systems in one-\-dimensional many
body physics. We derive the exact wave function for the
Thirring/Luttinger and Coset fermion models and apply our results to
derive the universal behavior of the wave functions of the Heisenberg
antiferromagnets
and of the Sutherland model. We find explicit forms for the wave
functions in the density and in the Grassmann representations. We show
that these wave functions always have the Jastrow factorized form and
calculated the exponent. Our results agree with the exponents derived
from the Bethe Ansatz for the Sutherland model and the Haldane-\-Shastri
spin chain but apply to all the systems in the same universality class.}

\body
\vbox to10truemm{}
\endtopmatter

\def\sch{Schr\"odinger}
\def\heis{Heisenberg}

\head{I. Introduction}

The quantization of physical systems with a
finite number of degrees of freedom is usually carried out in either the
\sch\ picture or in the \heis\ picture. In Quantum Field Theory (QFT) and
in Many-\-Body Physics the most direct approach involves the use of
second quantization and the \heis\ picture is natural. In this approach,
the theory focuses on the construction of the ground state and on the
computation of the Green functions. The \sch\ picture of a QFT, and of a
generic many body system, is usually very complicated. The wave
functions, whose properties can be guessed by intuition in simple
systems with a finite number of degrees of freedom, become very complex
objects. In the conventional second quantized picture, the states of
the Fock space are represented in terms of eigenstates of the number
operators. In the \sch\ approach, the states are wave {\it functionals}
of the field configurations. These functionals are very difficult to
define\refto{symanzik} and they are very rarely used.

In most cases,
these theories are discussed in the context of some form of
perturbation theory. The unperturbed states are usually very simple and
a Feynman diagrammatic approach is certainly simplest in the
\heis\ picture. The situation becomes radically different when it
comes to the
study of the non-\-perturbative structure of these theories. To date,
there are only two approaches to non-\-perturbative physics. One
are the semiclassical and mean-\-field approaches. The other is the
explicit construction of the exact or approximate eigenstates of the
systems. While the semiclassical approach is usually presented in terms
of path-\-integrals, the eigenstate approach uses the first quantized or
\sch\ picture. Although very rarely used in QFT, it is the conventional
approach for many-\-body physics, \eg\ the BCS approach to
superconductivity and the Laughlin picture of the Fractional Quantum
Hall Effect. Although much has been gained from the
semiclassical approach\refto{rajaraman}, much of what is known about
non-\-linear theories in 1+1 dimensions is due to the Bethe Ansatz
approach\refto{leshouches}. However, unlike the BCS and Laughlin
wavefunctions, the Bethe Ansatz wavefunctions are so complex that their
structure is almost never discussed, except for some very special
systems\refto{sutherland,shastri,haldaner^2} for which the wave
functions become very simple.

In this paper we study the structure of the wave functionals of
fermionic theories in 1+1 dimensions using field-\-theoretic path
integral methods and extract their universal content. We focus our
attention on a number of field-\-theoretic model and on their
application to the physics of one-\-dimensional strongly correlated
systems and antiferromagnets. In particular, we consider
self-\-interacting theories of massless fermions, including free
fermions, the Thirring/Luttinger model and Constrained Fermi
systems (Coset models) and apply our results to the study of quantum
\heis\ antiferromagnets and to the Sutherland model. These systems have
the interest since they are
at a fixed point (\ie\ they exhibit conformal invariance), their Green
functions exhibit well known universal properties.

We use the
methods of references [\cite{anawave}] and [\cite{wavefun}] to show
that the ground state wave functionals of of the systems are
determined completely by
the knowledge of the Green functions. We introduce two representation of
the wave functions: the {\it density} and the {\it Grassmann}
representations. The density representation, which is closely related to
the conventional particle number representation of second quantization,
is given in terms of the density-\-density correlation functions. These
correlation functions are exactly calculable by path integral methods in
all the systems that we discuss which exhibit conformal and
Kac-\-Moody invariance. The wave functionals  have universal properties
which are determined by the universal properties of the Green functions.
The Grassmann representation gives a more detailed picture of the wave
function in terms of fermion coherent states.

The results presented here are directly applicable to a number of
interesting models of strongly correlated systems in one space
dimension. For instance, this
result allows us to show that the wave functions of the Sutherland model
have the same universal properties as those of the Thirring/Luttinger
model. This is interesting since the Sutherland model can be
viewed as a
non-\-relativistic {\it regularization} of the Thirring/Luttinger
model and it is solvable by the methods of the Bethe
Ansatz\refto{sutherland}. The application of the methods of the Bethe
Ansatz to models of massless relativistic fermions, such as the
Thirring/Luttinger model, is plagued with difficulties. Our results on
Coset models are directly applicable to quantum antiferromagnets. It is
worthwhile to stress that while the Bethe Ansatz wave functions are
usually so complex that there structure is never studied, they too have
universal features which can be determined directly with our methods.

The paper is organized as follows. In section II we show how to extract
the wave functionals from the path integral formulation of field
theory. We begin with
a careful definition of the Hilbert space. Two descriptions of the wave
functionals are introduced: a) the second quantized or ``density"
picture and b) the Grassmann or Bargmann picture. In section III we
illustrate these ideas in the context of free fermions and free bosons.
These results had been derived previously using
\sch\ methods\refto{wheeler,symanzik,luscher,jackiw}. In section IV we
apply
these methods to the massless Thirring/Luttinger model. Here we give an
explicit expression for the asymptotic behavior of the wave functional
in both the density and Grassmann representations. Our results indicate
that, while the Grassmann representation is a sum of terms which exhibit
left-\-right factorization, the density representation does not. In
addition both pictures yield wave functionals with a Jastrow form but
with different exponents. In section V we
discuss Coset models and derive the wave functionals for these systems.
In section VI we discuss the connection between the
\heis\ and Sutherland models ( and their generalizations) with the
Thirring/Luttinger model
and show that the wave functions have the same long distance behavior.
The connection between the \heis\ model and the Coset field theories is
used to derive the long distance form of the wave functions. We show
that the asymptotic form at long distances of the wave function for the
nearest-\-neighbor and the $1/|r|^2$ models have the same
Jastrow exponent
(as required by universality). The conclusions are drawn in section VII.

\head{II. Wave Functionals and Path Integrals}

Let us begin by introducing a Hilbert space adequate to describe the
states of the fermionic systems of interest. Our description is equally
applicable to both relativistic and non-\-relativistic systems in the
second quantized formulation. In either case the space of states is
Fock space. We will introduce two representations of the wave
functions: a) the density representation and b) the Grassmann
representation.

Let $\ket{0}$ be a reference state on
which we will build the physical Hilbert space. In general we will
assume that $\ket{0}$ is a filled fermi sea, although in
non-\-relativistic theories is sometimes useful to consider the case in
which $\ket{0}$ is the empty state (\ie\ the state with no particles).
For the case of case of non-\-relativistic fermions we will
denote with ${\hat c}^{\dagger}(x)$ and ${\hat c}(x)$ a set of
creation and annihilation operators. Then a basis for the $N$-\-particle
subspace of Fock space has the form $\ket{x_1,\ldots,x_N}={\hat
c}^{\dagger}(x_1) \ldots {\hat c}^{\dagger}(x_N) \ket{0}$. These states
are eigenstates of the particle density operator ${\hat \rho}(x)= {\hat
c}^{\dagger}(x) {\hat c}(x)$ with eigenvalue $\rho(x)=\sum_{j=1}^N
\de(x-x_j)$. We will call this the {\it density} representation and
label these states by the density eigenvalue, \ie\
$\ket{x_1,\ldots,x_N}\equiv\ket{\rho(x)}$. For the case of dense
systems, we will label the states by their density eigenvalue in the
form $\ket{[\rho]}$, where $\rho(x)$, the eigenvalue of ${\hat
\rho}(x)$, is a distribution.

Now,
the density eigenstates are antisymmetric under particle exchange.
However, the density
eigenvalue is {\it even} under exchange. Thus, the density eigenvalue
only specifies the presence or absence of a particle at a given point
but says nothing about their relative ordering. Nevertheless, we will
continue to denote the states by the density eigenvalue and assume that
an antisymmetrization (or symmetrization, for the case of bosons) has
been adopted. Notice that, in spite of these caveats, the states
represented by the density eigenvalue still satisfy the exclusion
principle.

There is an
alternative representation in which the antisymmetry of the states is
apparent. In this representation, the coordinates which label the states
are Grassmann variables. We will refer to this as the {\it Grassmann}
representation and it
is the representation in terms of fermion coherent states\refto{jackiw}.
Let
$\xi_1,\ldots,\xi_N$ be a set of $N$ Grassmann variables. The coherent
state is defined to be
$\ket{\xi_1,\ldots,\xi_N}=\exp\{\sum_{j=1}^{N}\xi_j {\hat
c}^{\dagger}(x_j)\}\ket{0}$. These states are overcomplete.
Let $\ket{\Psi}$ be an arbitrary state of the Hilbert space. The wave
functions are inner products between the states and a set of reference
states. In the density representation the wave functions are the usual
antisymmetric ``orbital" wave functions
$\Psi(x_1,\ldots,x_N)=\ipr{x_1,\ldots,x_N}{\Psi}$.
In the Grassmann representation, they are Grassmann polynomials
$\Psi(\xi_1,\ldots,\xi_N)$. The orbital wave functions are the
number coefficients of the Grassmann polynomials.

We will now derive a set of expressions for the wave functions, for
both the
density and Grassmann representations, in terms of the correlation
functions of the specific system. For simplicity, we will discuss
only the ground state wave function. These formulas generalize without
difficulty for excited states. For the most part they also apply almost
without changes to both relativistic and non-\-relativistic systems.
We will discuss first the case of systems of non-\-relativistic
fermions and then generalize our results for relativistic systems.

We begin with the density representation.
Following ideas that go back to Landau's work on superfluid
helium\refto{landauhelium}, Dashen and  Sharp \refto{dashen} (DS) have
shown that density
representation follows naturally from the symmetries of the physical
system. If the Hamiltonian has a conserved charge, such as particle
number, then the density ${\hat \rho}(x)={\hat c}^{\dagger}(x) {\hat
c}(x)$ and the current ${\hat j}(x)=\frac{1}{2i}[{\hat
c}^{\dagger}(x)\pd_x {\hat c}(x)-{\hat c}(x)\pd_x {\hat
c}^{\dagger}(x)] $ are the time and space components of a
locally conserved vector which obeys the continuity equation,
$\pd_t {\hat \rho}(x,t)+\pd_x {\hat j}(x,t)=0$. From the
equal-\-time canonical anticommutation relations of the field operators,
it is easy to see that the density and current operators must satisfy
the equal-\-time commutation relations
$$
[{\hat \rho}(x),{\hat j}(x')]=-i \pd_x\left(\de(x-x'){\hat
\rho(x)}\right)\eqno(2.1)
$$
and
$$
[{\hat \rho}(x),{\hat \rho}(x')]=[{\hat j}(x),{\hat j}(x')]=0\eqno(2.2)
$$
Notice that these commutation relations are non-\-trivial only for {\it
dense} systems, \ie\ systems for which the local density operator has
non-\-vanishing expectation values. In other words, this algebra only
applies to operators acting on a Hilbert space of states obtained by
acting finitely on a reference state which represents a {\it filled
Fermi sea}. We will come back to this point when we discuss relativistic
systems.

These commutation relations imply that the states can be either
eigenstates of the density {\it or} eigenstates of the current but not
of both. Dashen and Sharp further showed that $\rho(x)$ and $j(x)$ are a
satisfactory set of coordinates in the sense that the set of states
labelled by $\rho(x)$ {\it or} by $j(x)$ span the Hilbert space. In the
{\it density} representation, the current operators are represented by
$$
{\hat j}(x) \ket{\Psi[\rho]}\equiv
-i \rho(x) \pd_x \left(\frac{\de}{\de
\rho(x)} \ket{\Psi[\rho]}\right)\eqno(2.3)
$$
(Recall, however, the caveats concerning antisymmetry). Since these
states form a complete basis of the Hilbert space, it
is possible to resolve the identity operator ${\hat I}$ in the form
$$
{\hat I}=\int \cD \rho \; \ket{[\rho]} \bra{[\rho]} \eqno(2.4)
$$
up to a normalization constant (which we absorb in the measure).

We will now follow the methods of references [\cite{anawave,wavefun}] to
write the wave functions in the density representation.
We begin by recalling that the absolute value squared of the ground
state wave function in the density representation $|\Psi_0[\rho]|^2$ is
given by
$$
\eqalign{
|\Psi_0[\rho]|^2=
\int \cD A_0\;& \exp\{-i\int dx_1
\;A_0(x_1)\;\rho(x_1)\}\cr&\lim_{A_0(x)\to A_0(x_1)\de(x_0)}\me{0}{T
\exp\{i\int d^2x\;A_0(x)\;{\hat j}_0(x)\}}{0}\cr}\eqno(2.5)
$$
where ${\hat j}_0(x)\equiv{\hat \rho}(x)$. This identity can be proven
by
inserting the decomposition of the identity, eq~(2.4), inside the vacuum
expectation value in the integrand of eq~(2.5) and by carrying out the
Fourier transforms. Please note that the Fourier transform is a path
integral at a fixed time $x_0=0$. The operators in this
expression are \heis\ operators of the system in the absence of sources.
The vacuum expectation value in the integrand of eq~(2.5) can be
calculated from the generating functional of density correlation
functions. Let us denote by $\cZ[A_{\mu}]$ the generating functional for
density and current correlation functions. $\cZ[A_{\mu}]$ is obtained
from the path integral of the system by coupling it minimally to an
external, classical, gauge potential $A_{\mu}$,\ie
$$
\cZ[A_{\mu}]=\int \cD {\bar \psi} \cD \psi\; \exp\{i S[{\bar \psi},
\psi, A_{\mu}]\}\eqno(2.6)
$$
Clearly, we have
$$
|\Psi_0[\rho]|^2=\int \cD A_0\; \exp\{-i\int dx_1 \;A_0(x_1)\;\rho(x_1)\}
 \lim_{A_0(x)\to A_0(x_1)\de(x_0)} \; \cZ[A_0,A_1=0]
\eqno(2.7)
$$
Eq~(2.7) tells us that $|\Psi_0[\rho]|^2$  is determined by the
generating functional of equal-\-time density correlation functions. In
the next sections we will make extensive use of eq~(2.7). This formula
was derived in reference [\cite{anawave}].

It is also possible to find an expression for the wave function itself,
not just for the absolute value squared. This is done by using a
generalization of the identity eq~(2.5). In order to do that we need
to be able to get information about the phase of the wave function,
not just of its absolute value squared. The commutation relations
eq~(2.2) imply that the phase information must be related to the current
operators. This leads us to consider the full generating function of
density and current correlation functions $\cZ[A_{\mu}]$. In particular,
we will consider the case in which the space and time components of
$A_{\mu}$ are non-\-zero on {\it different} time surfaces and look at
vacuum expectation values of the form
$$
\cZ[A_0,A_1]\equiv
\me{0}{T \exp\{i\int d^2x\;A_1(x)\;{\hat j}_1(x)\}\;
\exp\{i\int d^2x\;A_0(x)\;{\hat j}_0(x)\}}{0}
\eqno(2.8)
$$
where we must set $A_0(x)\to A_0(x_1)\de(x_0-t)$ first, $A_1(x)\to
A_1(x_1)\de(x_0-t')$ later and, finally, let $t'\to t$. In what follows
we will assume that these operations are done in the order that we have
just specified. This prescription amounts to a choice of order among the
operators.
By inserting now the resolution of the identity eq~(2.4) inside the
matrix
element in eq~(2.8) and after recalling the expression for the current
operator in
the density representation, eq~(2.3), we can write eq~(2.8) in the form
$$
\eqalign{
\cZ[A_0&,A_1]\equiv\cr
&\int \cD \rho
\;\exp\{\int dx_1\;\pd_x A_1(x_1) \rho(x) \frac{\de}{\de
\rho(x)}\} \Psi_0^*[\rho]\;\Psi_0[\rho]\; \exp\{i\int dx_1\;
A_0(x_1)\rho(x_1)\}\cr &= \int \cD \rho\;  \Psi_0^*[\rho+\rho
\pd_xA_1]\;\Psi_0[\rho]\;\exp\{i\int dx_1\; A_0(x_1)\rho(x_1)\}\cr}
\eqno(2.9) $$
Hence, after a Fourier transform we can write
$$
\Psi_0^*[\rho+\rho \pd_xA_1]\;\Psi_0[\rho]=\int \cD A_0 \exp\{-i\int
dx_1\; \rho(x_1) A_0(x_1)\} \cZ[A_0,A_1]
\eqno(2.10)
$$
where $\cZ[A_0,A_1]$ is the expression calculated with the prescription
of eq~(2.8). We can now choose $A_1(x)$ to be such that, for arbitrary
$\rho(x)$, $\rho(x)+\rho(x) \pd_xA_1(x)={\bar \rho}$, where $\bar \rho$
is the average density. We can further pick the arbitrary choice of
setting $\Psi_0[\bar \rho]=1$ ( \ie\ a fixed real number). With these
prescriptions we can now write an expression for the full wavefunction:
$$
\Psi_0[\rho]=\int \cD A_0 \exp\{-i\int dx_1\; \rho(x_1) A_0(x_1)\}
\cZ[A_0,A_1] \eqno(2.11)
$$
provided that $A_1(x)$ satisfies the condition $\rho(x)+\rho(x)
\pd_xA_1(x)={\bar \rho}$ and that $\cZ$ is calculated in the limit (and
with the order) prescribed above.

The density representation can be applied almost without changes to
relativistic fermions. Let us consider a relativistic Fermi field
$\psi(x)$. For the most part, we will not write the two Dirac components
explicitly in what follows. The field $\psi$ obeys canonical equal
anticommutation relations. The $U(1)$ charge current  $J_{\mu}={\bar
\psi} \gamma_{\mu} \psi$, normal-\-ordered relative to the filled Dirac
sea, obeys the equal-\-time $U(1)$ Kac-\-Moody algebra
$$
[J_0(x),J_0(x')]=[J_1(x),J_1(x')]=0\qquad
[J_0(x),J_1(x')]=-\frac{i}{\pi}\pd_x\de(x-x')\eqno(2.12)
$$
These commutation relations, which are the relativistic version of the
non-\-relativistic commutation relations of eq~(2.2), determine the
choice of representation for the wave functions. Here too, we will adopt
the density representation, \ie\ we will pick the wave functions to be
eigenstates of the density operator $J_0(x)$. However, in general, the
density
eigenvalue alone does not specify the state completely. For instance,
for free fermions it is also
necessary to specify the densities of left and right movers separately.
In the case of interacting theories, because of the presence of the
chiral anomaly, it is not generally possible to conserve both the right
and left moving densities separately. One consequence of the anomaly is
the existence of anomalous dimensions of various operators. In what
follows we will assume that the regularization that we are using is such
that the $U(1)$ charge current is conserved even in the presence of
interactions.

We can now use the same approach that we developed above
for non-\-relativistic systems, and apply it to relativistic models,
such as the Thirring/Luttinger model. We will consider the Hilbert
space built on the state $\ket{0}$, the {\it filled Dirac sea}. The
$U(1)$ charge density is a diagonal operator in the basis of states
$\{\ket{x_1, x'_1;\ldots;x_n, x'_n}\}$ which denotes a set of $n$
particles (electrons) and $n$ antiparticles (holes), with the eigenvalue
$\rho(x)=\sum_{j=1}^n(\de(x-x_j) -\de(x-x'_j))$. The probability
$|\Psi_0[\rho]|^2$ that a given $U(1)$ charge density profile $\rho(x)$
is observed in the ground state is still given exactly by the
same equation that we derived for non-\-relativistic systems, eq~(2.7).
If we set the amplitude for the state with uniform (zero) density to
unity, the {\it amplitude}
$\Psi_0[\rho]$ is given by eq~(2.11) but with $A_1(x)$ given by
$$
A_1(x)=-\pi \int_{-\infty}^x dx' \rho(x')\eqno(2.13)
$$
This change is consequence of the fact that the Kac-\-Moody
algebra of eq~(2.12) is a little simpler that its non-\-relativistic
version eq~(2.2).

We end this section with a few comments on how the Grassmann
representation is constructed using these methods. Except for the
presence of various indices ( Dirac components, \etc), the method
is the same for both relativistic and non-\-relativistic theories.
The Grassmann representation of the wave functions can be constructed in
a manner very much analogous to the construction of the density
representation. The main difference is that now we want to project the
ground state onto  fermion coherent state $\ket{[\chi]}$. It is
straightforward to show that this can be achieved by considering the
generating function of {\it fermion} correlation functions $\cZ[{\bar
\eta},\eta]$, \ie
$$
\cZ[{\bar \eta},\eta]=\int \cD {\bar \psi} \cD \psi \exp\{iS({\bar
\psi},\psi)+i\int d^2x [{\bar \eta} \psi+{\bar \psi}\eta]\}\eqno(2.14)
$$
The probability for the state $\ket{[\chi]}$ to occur in the ground
state is
$$
|\Psi[{\bar \chi},\chi]|^2=\int \cD {\bar \eta} \cD \eta \cZ[{\bar
\eta},\eta] \;\exp\{-i\int dx [{\bar \eta} \chi+{\bar \chi}\eta]\}
\eqno(2.15)
$$
with the same equal-\-time limits used in the density representation.

\head{III. Wave Functionals for Free Theories in two dimensions}

As a warm-up exercise, we will consider first the cases of a free
relativistic scalar field ( a theory of bosons) and free relativistic
fermions, both in $1+1$-\-dimensions. The non-\-relativistic case can
also be
considered without difficulty ( in any case,  non-\-relativistic
fermions at finite density are, in one space dimension, equivalent to
relativistic fermions). For simplicity we will work in Euclidean space
\ie $ix_0\to\tau$. Note that besides the usual modifications in the partition
functions
$iS_{M}\to S_{E}$, the density-current correlation
functions (2.1), (2.12) acquires an extra factor of $i$.

\subhead{a. Free bosons}

The (Euclidean) action for a free scalar field is:
$$
S[\phi] = {1\over 2} \int d^2x \partial_{\mu}\phi \partial_{\mu}\phi
\eqno(3.1)
$$
so that the partition function $Z[J]$ is given by:
$$
Z[J] = \int D\phi \exp (-S[\phi] + \int d^2x J.\phi)
\eqno(3.2)
$$
or,
$$
Z[J] = Z[0] \exp [-{1\over 2} \int d^2x d^2y J(x)G(x,y)J(y)] \eqno(3.3)
$$
with $G(x,y)$ the 2-dimensional Laplacian Green function,
$$
G(x,y) = {1\over 2\pi} \log \vert x - y \vert
.\eqno(3.4)
$$
As explained in the precedent section, one should take for the source $J$:
$$
J = J(x_1)\delta(x_0)
.\eqno(3.5)
$$
Then, calling
$$
\hat G(x_1,y_1) = \lim_{x_0 \to y_0} G(x,y) \eqno(3.6)
$$
we easily find, applying formula (2.7):
$$
\vert \Psi [\varphi] \vert^2 = \int DJ \exp [-{1\over 2} \int dx_1
dy_1 J(x_1)\hat G(x_1,y_1) J(y_1) + i\int dx_1 J(x_1)\varphi (x_1)]
\eqno(3.7)
$$
or,
$$
\vert \Psi [\varphi] \vert^2 =  \exp [-2\int dx_1 dy_1 \varphi(x_1)
\hat G^{-1}(x_1,y_1) \varphi(y_1)].\eqno(3.8)
$$

We have now to evaluate $ \hat G^{-1}(x_1,y_1)$ satisfying:
$$
\int dz_1 \hat G^{-1}(x_1,z_1)\hat G(z_1,y_1) = \delta (x_1 - y_1)
.\eqno(3.9)
$$
Writing:
$$
\hat G^{-1}(x_1,y_1)  = \int {dp\over 2\pi} g^{-1}(p)\exp [ip(x_1-y_1)]
\eqno(3.10)
$$
one gets:
$$
g^{-1}(p) = -2 \vert p \vert
\eqno(3.11)
$$
or
$$
\hat G^{-1}(z) = -{2\over \pi} {1\over z^2}
\eqno(3.12)
$$
so that the ground-state wave-functional for two dimensional free-bosons
is given by:
$$
\vert \Psi [\varphi] \vert^2 = \exp [-{4\over \pi} \int dx_1 dy_1
\varphi (x) {1\over (x-y)^2} \varphi (y)] \eqno(3.13)
$$
which is the correct answer.

\subhead{b. Free fermions}

Let us now compare this result with that obtained for the ground-state wave
functional of a system of two-dimensional free fermions. The (Euclidean)
action for fermions
coupled to Grassmann valued external sources $\bar \eta , \eta$
is given by:
$$
S[\bar \eta , \eta] = \int d^2x (\bar \psi i\not\!\partial \psi  - \bar \eta
 \psi
- \bar \psi \eta ).\eqno(3.14)
$$
(With our conventions $\gamma_{\mu} \gamma_5 = i \epsilon_{\mu
\nu}\gamma_{\nu}$ and $\gamma_5 = i\gamma_0\gamma_1$).

The partition function $Z[\bar \eta, \eta]$ is then given by:
$$
Z[\bar \eta , \eta] = \int D\bar \psi D\psi \exp (S[\bar \eta , \eta])
 \eqno(3.15)
$$
or
$$
Z[\bar \eta , \eta] = \exp (-\int d^2x d^2y \bar \eta (x) G_F(x,y) \eta (y))
\eqno(3.16)
$$
where the fermionic Green function $G_F$ given by:
$$
G_F(x-y) = -{i\over 2\pi} {\gamma_{\mu}(x_{\mu} - y_{\mu})\over (x-y)^2}
\eqno(3.17)
$$
As before, we take $\eta(x) = \eta(x_1)\delta(x_0)$
(and the same for $\bar \eta $) and then we need to compute:
$$
\hat G_F(x_1,y_1) = \lim_{y_0 \to x_0} G_F(x-y) = -{i\over 2\pi}
{\gamma_1\over (x_1-y_1)} \eqno(3.18)
$$
Then, applying formula (2.15), we have:
$$
\vert \Psi[\bar \chi, \chi] \vert^2 = \int D\bar \eta D\eta
Z[\bar \eta , \eta] \exp(-i\int d^2x (\bar \chi \eta +\bar \eta \chi ))
\eqno(3.19)
$$
or:
$$
\eqalign{
\vert \Psi[\bar \chi, \chi] \vert^2 & =  \int D\bar \eta D\eta
\exp(-\int dx_1dy_1 \bar \eta (x_1) \hat G_F(x_1,y_1) \eta (y_1) \cr
&-i\int dx_1 (\bar \chi \eta +\bar \eta \chi )).\cr}
\eqno(3.20)
$$
The quadratic path-integral in (3.20) can be easily evaluated.
The answer is:
$$
\vert \Psi[\bar \chi, \chi] \vert^2  = \exp ( \int dx_1 dy_1 \bar
\chi(x_1) \hat G_F^{-1}(x_1,y_1) \chi(y_1)) \eqno(3.21)
$$
where $\hat G_F^{-1}(x_1,y_1)$ satisfies:
$$
\int dz_1 \hat G_F^{-1}(x_1,z_1) G_F(z_1,y_1) = \delta(x_1 - y_1).
\eqno(3.22)
$$
Fourier transforming (3.22) we get,
$$
\hat G_F^{-1}(z) = \int {dp\over 2\pi} g(p)\exp (ipz)
\eqno(3.23)
$$
with
$$
g(p) = -2 \gamma_1 {p\over \vert p \vert }.\eqno(3.24)
$$

We then finally have:
$$
\vert \Psi[\bar \chi, \chi] \vert^2 = \exp (-{2i\over \pi} \int dx_1 dy_1
\bar \chi(x_1) {\gamma_1\over x_1 - y_1} \chi(y_1)).
\eqno(3.25)
$$
This wave functional can be written in a more familiar way if work in a
subspace of the Hilbert space with a fixed number of particles. Indeed
expanding the exponential in the Grassmann variables we get
$$
\vert \Psi[\bar \chi, \chi] \vert^2 =  \sum_{n=0}^{\infty}
\left({-2i\over \pi}
\right)^n {1\over n!^2}\int\left(\prod_{i=1}^n dx_idy_i\right)
\prod_{j=1}^n\left({\bar \chi}_{\alpha_j}(x_j)\chi_{\beta_j}(y_j)\right)
{}~F_0[\{x_j,y_j\}]_{\{\alpha_k \beta_k\}}
\eqno(3.26)
$$
where $j,k=1,\ldots, n$ and the function $F_0[x_1,y_1;...]_{\alpha_1
\beta_1
\ldots \alpha_n \beta_n}$ is given by
$$F_0[x_1,y_1;...;x_n,y_n]_{\alpha_1 \beta_1 ... \alpha_n
\beta_n}=(\gamma_1)_{\alpha_1 \beta_1}...(\gamma_n)_{\alpha_n \beta_n}\times
\det{1\over (x_i-y_j)}\eqno(3.27)$$
which is the wave function of a system of $n$ particles and $n$ antiparticles.

We have to mention here that the wave function (3.25) is the same
we obtain working canonically  in the Schrodinger representation.
In this approach the Dirac fermions $\bar \psi$ and $\psi$ are represented by
differential operators in the Grassmann variables ${\bar \chi}$ and
$\chi$ acting over the wave functional (3.25)\refto{jackiw}
$${\bar \psi}_i={1\over \sqrt{2}}\left({\bar \chi}_i +(\gamma_0)_{j i}{\delta
\over \delta \chi_j}\right)\eqno(3.28)$$
$$\psi_i={1\over \sqrt{2}}\left(\chi_i +(\gamma_0)_{i j}{\delta
\over \delta {\bar \chi}_j}\right).\eqno(3.29)$$
They satisfy the canonical commutation relations
 $$\{ \psi(x)_i , {\bar \psi}(y)_j\}=(\gamma_0)_{i j}
\delta(x-y).\eqno(3.30)$$

Using the property (3.22) (\ie $G_F^2 \propto I$) we can easily verify
that the
wave functional (3.25) is annihilated by the second quantized hamiltonian
$$
\int d^2x {\bar \psi}(x){\partial \psi (x)\over \partial x}~\Psi[{\bar
\chi},\chi]=0.\eqno(3.31)
$$

It might be worthwhile to reanalize the case of free fermions but
considering
that external sources couple to currents and not directly to fermion
fields.
What we have in mind is the formulation of two-dimensional fermionic
models
\` a la Sugawara, that is, in terms of currents $j_{\mu} = \bar \psi
\gamma_{\mu} \psi $ so that, in our
quantum mechanical language wave functionals, after a polarization choice,
should depend say on $j_0$. Then, instead of action (3.14) we start
from:
$$
S[A_{\mu}] = \int d^2x (\bar \psi i\not\!\partial \psi  - \bar \psi
\gamma_{\mu} \psi A_{\mu} ) \eqno(3.32)
$$
The corresponding partition function is, after path-integration\refto{fidel}:
$$
Z[A] = \exp (-{1\over 2\pi} \int d^2x d^2y A_{\mu}({\vec x}) D_{\mu
\nu}({\vec x},{\vec y}) A_{\nu}({\vec y}) ) \eqno(3.33)
$$
with:
$$
D_{\mu \nu}({\vec x}-{\vec y}) = ( \delta_{\mu \nu} - \partial_{\mu}
\bigtriangledown ^{-1}\partial_{\nu}) \delta^{2}({\vec x}-{\vec y})
\eqno(3.34)
$$
and
$$
\bigtriangledown^{-1}({\vec x},{\vec y}) = -{1\over 2\pi} \ln \vert
{\vec x} - {\vec y} \vert \eqno(3.35)
$$

Following the prescription of section II we can now
compute the density-representation  ground-state wave-functional.
We get:
$$
\vert \Psi_0[\rho] \vert^2 = \exp\left(-{\pi\over 2}\int dx_1dy_1
\rho (x_1)\hat D_{00}^{-1}(x_1,y_1)\rho (y_1)\right)
{}.
\eqno(3.36)
$$

The calculation of $\hat D_{00}^{-1}$ can be done more easily in
momentum space.
In terms of the Fourier transform of the vector source $A_{\mu}(x)$,
$$
A_{\mu}(x)=\int {dp\over 2\pi} {\tilde A}_{\mu}(p) \exp (ipx)\eqno(3.37)
$$
the partition function (3.33) takes the form
$$Z[A] = \exp (-{1\over 2\pi} \int d^2p {\tilde A}_{\mu}(p) {\tilde
D}_{\mu \nu}( {\tilde A}_{\nu}(-p) )
\eqno(3.38)
$$
where
$$
{\tilde D}_{\mu \nu}(p)
=\delta_{\mu \nu}-{p_{\mu}p_{\nu}\over {\vec p}^2}\eqno(3.39)
$$
is the Fourier transform of the current-current correlation function
(3.34). We
are interested in the equal time correlation function of densities
which is obtained integrating ${\tilde D}_{0 0}$ over the frequencies
$$
 \eqalign{<\rho(p)\rho(-p)>|_{\Delta t=0}&=\int dp_0 {1\over
\pi}\left(1- {p_0^2\over p^2+p_0^2}\right)\cr
&=|p|\cr}.\eqno(3.40)
$$
Hence the wave function (3.36) can be written as
$$
 \vert \Psi_0[\rho] \vert^2 = \exp(\pi\int dp \rho(p){1\over |p|}
\rho(-p)) . \eqno(3.41)
$$
To return to space-time variables we use the following identity
$$
{1\over |p|}=\int {d\omega\over \pi}{1\over \omega^2+p^2}.\eqno(3.42)
$$
in order to write
the Fourier anti-transform of the kernel ${1\over |p|}$ as
$$
{\cF}^{-1}\left[{1\over |p|}\right]=2\int {d\omega dp\over (2\pi)^2}
{1\over \omega^2+p^2}e^{i(\omega t + p x)}|_{t=0}\eqno(3.43)
$$
which is nothing but the propagator of a two dimensional massless bosonic
theory. Using the well-known result ${1\over \pi}\ln \mu |x|$ for
(3.43) we can finally write for the wave function (3.41) in real space
$$
\vert \Psi_0[\rho] \vert^2 = e^{\int dxdy
\rho (x)\ln\mu |x-y| \rho (y)}.
\eqno(3.44)
$$
We can now find the probability of any particular density profile. For
instance, we can look for the probability of a configuration in which
the $N$ particles are located at coordinates $x_1,\ldots, x_N$ and N
antiparticles (holes) are located at $y_1,\ldots, y_N$ ( regardless of
whether they are left or right movers ). For such configurations, the
density profile is $\rho(x)=\sum_{j=1}^n(\de(x-x_j) -\de(x-y_j))$. By
substituting this density profile in eq~(3.44) we find the result
$$
|\Psi_0(\{x_j\};\{y_j\})|^2=\cN
\frac{\prod_{i<j}|x_i-x_j|^{2}|y_i-y_j|^{2}}
{\prod_{i,j}|x_i-y_j|^{2}}\eqno(3.45)
$$

Eq~(3.45) gives the probability, not the amplitude, for this
configuration. We will
show below that, up to an overall sign, the full wave function is just
the square root of this expression.
This can be verified much in the same way as we did for the
Grassmann
representation: the wave function associated with eq~(3.44) is
annihilated by the Hamiltonian. In order
to prove this statement we must write the Hamiltonian in the density
representation. Of course this is the Sugawara-Sommerfield form of the
Hamiltonian which reads
$$
\cH=2\pi\int d^2x :\rho(x)^2+j(x)^2:\eqno(3.46)
$$
where the dots denotes normal ordering. In eq~(3.46) we have restored a
factor of $i$ to the definition of the current $j(x)$, which is now
a manifestly hermitian operator. Looking the equal time commutator
$$
[\rho(x),j(y)]=-i{1\over \pi}\partial_{x} \delta(x-y).\eqno(3.47)
$$
we see that the current $j(x)$ can be represented as a functional
differential operator in the variable $\rho(x)$ by
$$
j(x)=-i{1\over \pi}\partial_{x} {\delta\over \delta
\rho(x)}.\eqno(3.48)
$$
Then, the Hamiltonian takes the form
$$
\cH=2\pi\int d^2x~:\rho(x)^2+{1\over \pi^2}\left(\partial_{x}
{\delta\over \delta \rho(x)}\right)^2:\eqno(3.49)
$$
The action of the current over the wave function is given by
$$
\eqalign{j(x)\Psi_0[\rho]&=-i{1\over \pi}\partial_x \int dy
\ln\mu|x-y|\rho(y) \Psi_0[\rho]\cr
                           &=-i{1\over \pi}\int dy {1\over
x-y}\rho(y)\Psi_0[\rho]\cr}\eqno(3.50)
$$
showing that $j(x)$ acts over $\Psi_0[\rho]$ as the  Hilbert transform
of the density.
For the composite operator $:j(x)^2:$ we obtain
$$
:j(x)^2:\Psi_0[\rho]=-{1\over \pi^2}\int dy dz {1\over x-y}{1\over x-z}
\rho(y) \rho(z) \Psi_0[\rho]\eqno(3.51)
$$
and, using the property eq~(3.22), we find
$$
\int dx :j(x)^2: \Psi_0[\rho]=-\int dx \rho(x)^2
\Psi_0[\rho]\eqno(3.52)
$$
which proves our statement
$$
\cH \Psi_0[\rho]= 0.\eqno(3.53)
$$

\head{IV. The Thirring/Luttinger Model}

In this section we will compute the ground state wave functional for an
interacting fermionic model: the Thirring model. As we did with the free
fermions we can express the wave functional as a functional either of the
fermionic variables ${\bar \chi}, \chi$ or of the charge density
$\rho(x)$. We will start with the Grassmannian representation.

The Lagrangian for the Thirring model coupled to external sources
${\bar \eta}, \eta$ is

$$\cL=-{\bar \psi}i\slp \psi - {1\over 2} g^2 \left({\bar \psi}\gamma^{\mu}
\psi\right)^2 - {\bar \eta}\psi- {\bar \psi}\eta.\eqno(4.1)$$

We can eliminate the quartic term in (4.1) by making a
Hubbard-Stratonovich transformation. The generating
functional of the model becomes

$$Z[{\bar \eta},\eta]=\int D{\bar \psi}D\psi DA_{\mu}
e^{\int {\bar \psi}(i\slp+g\lnA)\psi - {1\over 2}A^2-{\bar \eta}\psi-
{\bar \psi}\eta~d^2x}.\eqno(4.2)$$

The integral over the fermionic variables is straightforward
$$
Z[{\bar \eta},\eta]=\int DA_{\mu}~\det \slD [A] e^{\int {\bar \eta}
(\slD [A])^{-1}\eta - {1\over 2}\int A^2}\eqno(4.3)
$$
where $(\slD [A])^{-1}$ is the inverse of the Dirac operator.
In order to perform the integration in the auxiliary field $A_{\mu}$
we use the following decomposition
$$
A_{\mu}={1\over g}\left(\partial_{\mu}\omega-\epsilon_{\mu \nu}
\pd_{\nu} \phi\right)\eqno(4.4)
$$
In terms of the fields $\phi$ and $\omega$ the Green function
$(\slD [A])^{-1}$ takes the simple form
$$
(\slD [A])^{-1}=e^{\gamma_5\phi+i\omega}|_{\vec x}
{i\over 2\pi} {\gamma_{\mu}(x_{\mu} - y_{\mu})\over ({\vec x}-{\vec
y})^2} e^{\gamma_5\phi-i\omega}|_{\vec y}.\eqno(4.5)
$$

For the determinant of the Dirac operator we use the result (3.33)-(3.34)
in terms of the fields $\phi$ and $\omega$. The partition function (4.3)
becomes
$$
Z[{\bar \eta},\eta]=\int D\phi D\omega e^{\int {\bar \eta}
(\slD [A])^{-1}\eta} e^{-{1\over 2\pi}S[\phi,\omega]}\eqno(4.6)
$$
where
$$
S[\phi,\omega]=\int~\{(1+{\pi\over g^2})\partial_{\mu}\phi
\partial_{\mu}\phi + {\pi\over g^2}\partial_{\mu}\omega
\partial_{\mu}\omega\}~ d^2x.\eqno(4.7)
$$
(The value of the fermionic determinant is given by equations
(3.33)-(3.34) up to a gauge-breaking term
$$
\alpha\int A^2d^2x\eqno(4.8)
$$
depending on the arbitrary regularization parameter $\alpha$. This means that
the quantization of the Thirring Lagrangian (4.1) leads to a one parameter
family of quantum theories. Our election $\alpha=0$, corresponding to a
gauge invariant definition of the fermionic determinant, is justified by the
fact that this value reproduces the standard solutions of the Thirring model).

Finally if we expand the exponential
term containing the Grassmann variables, we can perform the bosonic integrals
term by term. Instead of do this is more convenient to use first the
formula (3.3) for the square of the wave functional and leave the bosonic
integral to the end. Hence the square of the wave functional can be written as
$$
\vert \Psi[\bar \chi, \chi] \vert^2 = \int D\phi D\omega D\bar \eta D\eta
e^{\int {\bar \eta}  (\slD [A])^{-1}|_{\Delta t=0})\eta}~ e^{-{1\over
2\pi}S[\phi,\omega]}~e^{-i\int (\bar \chi \eta +\bar \eta \chi)}
\eqno(4.9)
$$
where
$$
(\slD [A])^{-1}|_{\Delta t=0}= e^{\gamma_5\phi(x,t)+i\omega(x,t)}
{i\over 2\pi}{\gamma_{1}\over (x-y)}
e^{\gamma_5\phi(y,t)-i\omega(y,t)}\eqno(4.10)
$$
is the equal time propagator.

The integration in the sources ${\bar \eta},\eta$ is Gaussian and
gives the result
$$
\vert \Psi[\bar \chi, \chi] \vert^2 = \int D\phi D\omega
e^{-\int {\bar \chi}  ((\slD [A])^{-1}|_{\Delta t=0})^{-1}
\chi}~e^{-{1\over 2\pi}S[\phi,\omega]}\eqno(4.11)
$$
(we have omitted in (4.11) a factor $\det[(\slD [A])^{-1}|_{\Delta
t=0}]$. It can
be proved, using for example a coherent-state definition of the functional
integral, that this factor is constant and does not play any role in our
problem).

The kernel $((\slD [A])^{-1}|_{\Delta t=0})^{-1}$
is computed identically to the one of the free fermion case giving the result
$$
((\slD [A])^{-1}|_{\Delta t=0})^{-1}=e^{-i(\gamma_5\phi(x,t)-\omega(x,t))}
{2i\over \pi}{\gamma_{1}\over (x-y)}
e^{-i(\gamma_5\phi(y,t)+\omega(y,t))}.\eqno(4.12)
$$
Finally the integration over the fields $\phi$ and $\omega$ can be done term by
term in the expansion of the exponential of the Grassmann variables and we have
for the wave functional
$$
\vert \Psi[\bar \chi, \chi] \vert^2 = \sum_{n=0}^{\infty} \left({2i\over \pi}
\right)^n {1\over n!^2}\int\left(\prod_{i=1}^n dx_idy_i\right)
\prod_{j=1}^n\left({\bar \chi}_{\alpha_j}(x_j)\chi_{\beta_j}(y_j)\right)
F[\{x_j,y_j\}]_{\{\alpha_k \beta_k\}}
\eqno(4.13)
$$
where the indices $\alpha_i,~\beta_i$ label the Dirac components and
$j,k=1,\ldots,n$. The function $F[x_1,y_1;...]_{\alpha_1 \beta_1 ...
\alpha_n
\beta_n}$ has two contributions: a factor corresponding to the free fermions
contributions (the function $F_0$ of equations (3.26) and (3.27)) and an
equal-time correlation function of vertex operators
with the bosonic action (4.7), coming from the interaction.
$$
\eqalign{
F[\{x_j,y_j\}&]_{\{\alpha_k \beta_k\}} =F_0[\{x_j,y_j\}]_{\{\alpha_k
\beta_k\}}  \times                           \cr
&\langle \exp(\sum_{j=1}^n s_j\phi(x_j,t))\;
\exp(\sum_{j=1}^n t_j\phi(y_j,t)) \rangle\;
\langle\exp(-i\sum_{j=1}^n\omega(x_j,t))
\exp(i\sum_{j=1}^n\omega(y_j,t) ) \rangle    \cr}
\eqno(4.14)
$$
where $j,k=1,\ldots,n$.
The variables $s_i~(t_i)$ are equal to $1$ if $\alpha_i~(\beta_i)=1$ and equal
to $-1$ if $\alpha_i~(\beta_i)=2$. The correlation function (4.14) can be
evaluated using the Wick theorem with the bosonic propagators
$$
\me{0}{\phi({\vec x})\phi({\vec y})}{0}=-{1\over 2(1+{\pi\over g^2})}\ln
|{\vec x}-{\vec y}|\eqno(4.15)
$$
and
$$
\me{0}{\omega({\vec x})\omega({\vec y})}{0}=-{g^2\over 2\pi}\ln |{\vec
x}-{\vec y}|.\eqno(4.16)
$$
For example, the term $F[x_1,y_1;...;x_n,y_n]_{1 2...1 2}$ is given by
$$
F[x_1,y_1;...;x_n,y_n]_{1 2...1 2}=[(\gamma_1)_{1 2}]^n\det {1\over
(x_i-y_j)}~{\prod_{i<j}|x_i-x_j|^{\mu}\prod_{i<j}|y_i-y_j|^{\mu}\over
\prod_{i,j}|x_i-y_j|^{\mu}}\eqno(4.17)
$$
where the exponent $\mu$ is
$$
\mu=-{1\over 2}{(g^2/\pi)^2\over 1+g^2/\pi}.\eqno(4.18)
$$

As we mentioned before we can also compute the wave functional of the Thirring
model in the density
representation. The first step is to obtain the generating functional of
current correlation functions of this model. Making a Hubbard-Stratonovich
transformation as we did in equation (4.2), the required generating functional
becomes
$$
\eqalign{Z[B_{\mu}]&=\int D{\bar \psi}D\psi DA_{\mu} e^{\int {\bar
\psi}(i\slp+g\lnA-\lnB)\psi - {1\over 2}A^2~d^2x}\cr &=\int DA_{\mu}
\det\left(i\slp+g\lnA-\lnB\right)~e^{-{1\over 2}\int A^2~d^2x}\cr}\eqno(4.19)
$$
where $B_{\mu}$ is the external source. After a shift in the integration
variables $A_{\mu}\to A_{\mu}-{1\over g}B_{\mu}$ and using the result
(3.33)-(3.34) for the fermionic determinant we get
$$
Z[B_{\mu}]=\int DA_{\mu}
e^{-{g^2\over 2\pi}\int d^2xd^2y~A_{\mu}({\vec x})D_{\mu \nu}({\vec
x},{\vec y})A_{\nu}({\vec y})}~e^{-{1\over 2}\int d^2x (A_{\mu}+{1\over
g}B_{\mu})^2}\eqno(4.20)
$$
with $D_{\mu \nu}({\vec x},{\vec y})$ defined in equation
(3.34). We can perform the Gaussian integration in $A_{\mu}$ and the partition
function takes the form
$$
Z[B_{\mu}]=e^{\int d^2xd^2y~B_{\mu}({\vec
x})\{{1\over 2g^2}{\tilde D}_{\mu \nu}^{-1}({\vec x},{\vec y}) - {1\over
2g^2}\delta_{\mu \nu}\delta^2({\vec x}-{\vec y})\}B_{\nu}({\vec y})}
\eqno(4.21)
$$
where ${\tilde D}_{\mu \nu}^{-1}({\vec x},{\vec y})$ is the Green function of
the operator
$$
\eqalign{{\tilde D}_{\mu \nu}({\vec x},{\vec y})&={g^2\over \pi}
D_{\mu \nu}({\vec x},{\vec y})+\delta_{\mu \nu} \delta^2({\vec x}-{\vec y})
\cr &=((1+{g^2\over \pi})\delta_{\mu \nu} - {g^2\over \pi}\partial_{\mu}
\bigtriangledown
^{-1}\partial_{\nu}) \delta^2({\vec x}-{\vec y}).\cr}\eqno(4.22)
$$
We can compute formally
$$
\eqalign{{\tilde D}_{\mu \nu}^{-1}&=\left(1+{g^2\over \pi}\right)^{-1}
{}~{1\over \delta_{\mu \nu} - \left(1+{g^2\over \pi}\right)^{-1}
\partial_{\mu} \bigtriangledown
^{-1}\partial_{\nu}}\cr
&=\left(1+{g^2\over \pi}\right)^{-1}~[\delta_{\mu \nu} +
\sum_{n=1}^{\infty} \left(1+{g^2\over \pi}\right)^{-n}
\left(\partial_{\mu} \bigtriangledown
^{-1}\partial_{\nu}\right)^n].\cr}\eqno(4.23)
$$
and using the fact that that the operator $\partial_{\mu} \bigtriangledown
^{-1}\partial_{\nu}$ is idempotent, we get for ${\tilde D}_{\mu \nu}^{-1}$
$$
{\tilde D}_{\mu \nu}^{-1}={1\over 1+{g^2\over \pi}}\left(
\delta_{\mu \nu} + {g^2\over \pi}\partial_{\mu} \bigtriangledown
^{-1}\partial_{\nu}\right) \delta^2({\vec x}-{\vec y}).\eqno(4.24)
$$
Finally adding the second term in the exponential of equation (4.21) we obtain
$$
Z[B_{\mu}] = \exp\left\{-{1\over 2\pi} {1\over 1+{g^2\over \pi}}\int
d^2x d^2y B_{\mu}({\vec x}) D_{\mu \nu}({\vec x},{\vec y})
B_{\nu}({\vec y})\right\}
\eqno(4.25)
$$
recovering the usual result for the
Thirring model: the current correlation functions are only modified (respect to
the free theory) by the presence of a constant factor. At this point is
straightforward to write the ground state wave functional following the same
steps we did in the free theory. The result is
$$
\vert \Psi_0[\rho] \vert^2 = \exp\left\{(1+{g^2\over \pi})\int dxdy \rho
(x)\ln\mu |x-y| \rho (y)\right\} \eqno(4.26)
$$

We can check this result by looking if the wave functional is
annihilated by the
Hamiltonian. For the Thirring model the Hamiltonian in the Sugawara-Sommerfield
form reads
$$
\cH=2\pi(1+{g^2\over \pi})\int d^2x :\rho(x)^2+j(x)^2:\eqno(4.27)
$$
where we have restored a factor of $i$ to the current $j(x)$. The
equal time commutator between the density and the current is
$$
[\rho(x),j(y)]=-i{1\over \pi}{1\over 1+{g^2\over \pi}}\partial_{x}
\delta(x-y).\eqno(4.28)
$$
and the current operator can be represented as
$$
j(x)=-i{1\over \pi}{1\over 1+{g^2\over \pi}}\partial_{x} {\delta\over
\delta \rho(x)}\eqno(4.29)
$$
It is easy to verify that, also in this case, the wave functional
of eq~(4.26) is a zero mode of the Hamiltonian
$$
\cH \Psi_0[\rho]= 0.\eqno(4.30)
$$

As a final test of our wave functional, we can verify the vacuum
expectation
current algebra in the Schrodinger representation. For example  for the
two point correlation function of currents we have
$$
\eqalign{<0|j(x)j(y)|0>&=-\int D\rho \Psi_0[\rho]{1\over (\pi+g^2)^2}
\partial_{ \rho(x)}\partial_{y} {\delta\over \delta
\rho(y)} \Psi_0[\rho]\cr
&=-{1\over \pi^2}\partial_{x}\partial_y\int dudv \ln|u-x| \ln|v_y|\int
D\rho \rho(u)\rho(v)|\Psi_0[\rho]|^2\cr
&={1\over 2\pi^2} {1\over 1+{g^2\over \pi}} {1\over
(x-y)^2}\cr}\eqno(4.31)
$$
which is the correct result.

A final comment about the wave functionals is in order. When we tested
our wave functionals looking if they were annihilated by the
hamiltonian we {\it assumed} that they were real (no phases appeared
when we
took the square root). Our assumptions were confirmed when we obtained
the right
results, equations (3.31), (3.52) and (4.30). However , in the
same spirit of the method developed in section I, it is possible to
obtain, for a wide class
of models, the {\it whole} ground state wave functional, \ie\ with the
phase
factors. Let us explain how does this idea work for the case of the
Thirring model.

We write generically the generating functional (4.19) as
$$
Z[B_{\mu}]=\int D{\bar \psi}D\psi \exp\left(-S[{\bar \psi},\psi]-
\int d^2x j_{\mu}B_{\mu}\right)\eqno(4.32)
$$
where $S[{\bar \psi},\psi]$ is the Thirring action.
We can write this equation as a vacuum expectation value
$$
Z[B_{\mu}]=\me{0}{\exp\{-\int d^2x j.B_1\}\exp\{-\int d^2x
\rho.B_0\}}{0}\eqno(4.33)
$$
and choose the fields $B_0$ and $B_1$ in the form
$$
B_0(x,t)=B_0(x)\delta(t-\tau)~~{\rm
and}~~B_1(x,t)=B_1(x)\delta(t-\tau).\eqno(4.34)
$$
Inserting in equation (4.34) the following resolution of the identity
$$
I=\int D{\hat \rho}~\ket{\hat \rho}\bra{\hat \rho}\eqno(4.35)
$$
we arrive to the result
$$
Z[B_{\mu}]=\int D{\hat \rho} \me{0}{\exp\{-\int d^2x j.B_1\}}{\hat
\rho} ~\exp\{-\int d^2x {\hat \rho}.B_0\}\Psi_0[{\hat \rho}].\eqno(4.36)
$$
Now we can use the representation (4.29) for the current in the density
representation and we have
$$
\exp\{-\int d^2x j(x)B_1(x)\}\ket{\hat \rho}=\ket{{\hat
\rho}+(\pi+g^2)^{-1}\partial_x B_1}.\eqno(4.37)
$$
Hence the partition function (4.33) takes the form
$$
Z[B_{\mu}]=\int D{\hat \rho}\; \Psi_0^*[{\hat \rho}+(\pi+g^2)^{-1}
\partial_x B_1] \Psi_0[{\hat \rho}] \exp\{-\int d^2x
{\hat \rho}.B_0\}\eqno(4.38)
$$
and leads to the quantity
$$\Psi_0^*[\rho+(\pi+g^2)^{-1}
\partial_x B_1] \Psi_0[\rho]= \int \cD B_0
Z[B_{\mu}] \exp\{\int d^2x
{\hat \rho}.B_0\}.\eqno(4.39)
$$
This last equation allows us to read-off the phase factor of the wave
functional $\Psi_0$.
Using equations (4.21)-(4.24) we can verify that for the Thirring model the
wave functional (4.26) is real.

\head{V. Coset Models}

As a last example of ground state wave functional of a two dimensional system,
we will consider a Conformal Coset Model. The fermionic version of the Coset
Model is obtained by projecting out a suitable subgroup $H$ of the original
fermionic group $G$. This projection is carried out, in the path-integral
approach, introducing Lagrange multipliers which forces the $H$-currents to
vanish. The Lagrange multiplier fields can be interpreted as gauge
fields. The total effect of the gauge  fields is to constrain the
fermion currents in $H$ to zero when acting on the physical
states\refto{goddard,polyakov,enri}.
The Coset Model $U(2)/SU(2)$
can be mapped onto the Heisenberg Model. We will exploit this connection
in Section VI.
The generalization to other coset of the form $U(N)/SU(n)$ is
straightforward.

By following the methods of the previous sections we construct the wave
functions by first choosing a reasonable set of labels. We will choose
to work in the density representation. Since the group $U(2)$ has
several generators, we must choose one ( or several of them). The
constraints impose the condition that all the $SU(2)$ generators
annihilate the physical states. For the coset $U(2)/SU(2)$, this leaves
only one $U(2)$ available generator, the ``charge" ${\bar \psi}
\gamma_0\psi$. In section VI we will show that this operator is
essentially the same as the
z-\-component of the spin in the Heisenberg model. Thus, we are led to
consider, once again, the generating functionals of the density
correlation functions. The Lagrangian of the coset fermion model is
given by
$$
\cL=\sum_{j=1}^2 {\bar \psi}^j i\slp \psi^j + \sum_{a=1}^3 \lambda_{\mu}^a
j_{\mu}^a\eqno(5.1)
$$
where $\psi^j~,~(j=1,2)$ are two Dirac fermions, $\lambda_{\mu}^a$ are Lagrange
multipliers and
$$
j_{\mu}^a=\sum_{i,j=1}^2 {\bar \psi}^i \gamma_{\mu} \sigma_{i j}^a
\psi^j\eqno(5.2)
$$
are the $SU(2)$ Noether currents (the $\sigma^a$ are the Pauli matrices).
These currents annihilate the physical states.

The generating functional of the correlation functions of the two
diagonal densities in $U(2)$ is given by
$$
\eqalign{
Z[a^1,a^2] &=\int D{\bar \psi}D\psi D\lambda\; \exp\left\{-\int d^2x \cL
- \int d^2x(\rho_1 a^1+\rho_2 a^2)\right\}\cr
&=\int D{\bar \psi}D\psi D\lambda\; \exp\left\{-\int d^2x{\bar
\psi}\gamma\cdot
(i\partial+\lambda+a^1 P_1 + a^2 P_2)\psi\right\}\cr}\eqno(5.3)
$$
where $P_1={I+\sigma^3\over 2},~~P_2={I-\sigma^3\over 2}$ and $\lambda_{\mu}=
\sum_{a=1}^3 \lambda_{\mu}^a\sigma^a$. We can eliminate in (5.3) the
$\sigma^3$
part of the coupling with the sources by a shift in the variable
$\lambda_{\mu}^3$. Then the partition function (5.3) can be written as
$$
\eqalign{Z[a^1,a^2]&=\int D\lambda \det\left(\gamma\cdot(i\partial+\lambda+
{a^1+a^2\over 2})\right)\cr
&=\exp\left\{-{1\over\pi}\int d^2x d^2y  {a^1(x)+a^2(x)\over 2}
D_{00}(x,y){a^1(y)+
a^2(y)\over 2}\right\} \int
D\lambda\;\exp\left\{-W[\lambda]\right\}\cr}\eqno(5.4)
$$
where $D_{00}(x,y)$ is given in equation (3.34) and $W[\lambda]$ is the
Wess-Zumino-Witten action\refto{polyakov2}.
Eq~(5.4) shows that the non-\-abelian structure factorizes and
decouples. This is a consequence of the simple structure of this coset.
In more complicated coset, such as those in the sequence of minimal
models, the non-\-abelian structure does not decouple.

The factorized structure of eq~(5.4) has important consequences for the
wave functions. Indeed, once the generating function is known, the
methods of the past sections
tell us that the wave functional of the $U(2)/SU(2)$ Coset Model is
$$
\eqalign{|\Psi[\rho_1,\rho_2]|^2
&=\int Da^1Da^2\;Z[a^1,a^2]\;\exp\left\{\int dx\;
(\rho_1a^1+\rho_2a^2)\right\}\cr
&=\int Da^1Da^2\; \exp\left\{-{1\over \pi}\int dx dy\; {a^1+a^2\over
2}D_{00}{a^1+a^2\over 2}\right\}\cr
&\qquad \exp\left\{\int dx \;\left[(\rho_1+\rho_2){a^1+a^2\over
2}+(\rho_1-\rho_2){a^1-a^2\over 2}\right]\right\}\cr
&=\delta [\rho_1-\rho_2]\;
\exp\left\{2\int dx dy\; \rho_1(x) \ln |x-y| \rho_1(y)\right\}
.\cr} \eqno(5.5)
$$
The factor $\delta [\rho_1-\rho_2]$ is a result of the constraint and it
simply means that the two densities have to be equal to each other. The
second factor implies that the wave function of the coset, as a
functional of one of the densities (say $\rho_1$) is equal to the {\it
square} of the wave function for free fermions ( see eq~(3.44)). Thus,
the Jastrow exponent $\nu$ for the $U(2)/SU(2)$ Coset Model is $\nu=2$.
In section VI we will make use of this result in the context of the spin
chains.

It is straightforward to generalize the result of eq~(5.5) for
the slightly more general case of the coset $U(N)/SU(n)$. In this case,
the constraint affects only $n-1$ diagonal generators of $U(N)$ (\ie\
those which span the Cartan subalgebra of $SU(n)$). If we define an
array of densities $\rho_1 ,\ldots, \rho_N$, the constraint forces the
first $n$ densities to be equal to each other and, say equal to
$\rho_1$, while the other $N-n$ densities remain unconstrained. The
wave function $\Psi_0[\rho_1 ,\ldots, \rho_N]$ becomes
$$
\eqalign{
|\Psi_0[\rho_1 ,\ldots, \rho_N]|^2&\equiv|\Psi_0[\rho_1
,\ldots,\rho_1,\rho_{n+1},\ldots \rho_N]|^2\cr
&=\exp\left\{n\int dx dy\; \rho_1(x) \ln |x-y| \rho_1(y)\right\}\cr
&\qquad \prod_{i=n+1}^N\;\exp\left\{\int dx dy\; \rho_i(x) \ln |x-y|
\rho_i(y)\right\}\cr}
\eqno(5.6)
$$
{}From this results we conclude that for all the simple coset the Jastrow
exponents are always integers. Non-\-integer exponents can only arise as
the consequence of the non-\-abelian structure.

\head{VI. Application to One-\-Dimensional Strongly Correlated
Fermi Systems and Antiferromagnets}

We will now discuss the applicability of the results of previous
sections to a number of models of one-\-dimensional many-\-body physics.
A number of these models are exactly solvable through the Bethe Ansatz.
These systems are said to be integrable and for this to happen a number
of conditions have to be satisfied. While these conditions do not
affect the low energy behavior, they do change the microscopic
properties of these systems. The Bethe Ansatz wave functions thus appear
to depend on a number of microscopic properties which are physically
irrelevant but are essential for integrability to hold. We will show in
this section that the wave functions of these systems also have
universal features. These universal properties, independent as they are
of the short distance physics, hold separately from the integrability
conditions. Most of the identifications of models listed below have been
known for some time. We present them here with the sole purpose of
making the connections clear.

There is a considerable number of systems in one-\-dimensional
many-\-body physics whose long distance behavior is governed by the
Thirring/Luttinger model, \ie\ they are in the same universality class.
The reason behind this universality has been known for a long time
\refto{lutherpeschel,emery,haldane}: if spin exchange is
ignored,
the low energy physics of all these systems reduces to the dynamics of
fermion excitations with momentum near the two Fermi
points (right and left movers) interacting \via\ backscattering
processes ( \ie\ processes which exchange left and
right movers). The basic assumption here is that, if the interactions
are not too strong, the only states which participate in the dynamics
are those which are sufficiently close to the Fermi ``surface". With
this assumptions, it is natural to separate the electron field into
its fast
and slow components. At weak coupling, the fast components can be
neglected (their only effect is to renormalize the coupling constant
and the energy scale for the slow components ). If we denote by
$\psi(x)$ the ``electron" field operator, the left
$\psi_L(x)$ and right $\psi_R(x)$ moving components of the Fermi field
are defined through the decomposition
$$
\psi(x)=e^{ik_Fx} \;\psi_R(x) + e^{-ik_Fx}\;  \psi_L(x)\eqno(6.1)
$$
and can be arranged in the form of a two-\-component spinor
$\psi_{\alpha}(x)$ ($\alpha=1,2$), with $\psi_1=\psi_R$ and
$\psi_2=\psi_L$.
At low energies the non-\-interacting dispersion curve can be linearized
in the vicinity of the Fermi points. The result is a spinor Fermi field
with an effective low energy Hamiltonian which
is identical to the Hamiltonian of
the relativistically invariant Thirring/Luttinger model. This
identification holds provided that (a) all energies are
rescaled so that the Fermi velocity is set to be one ,
$v_F=1$, and (b) all irrelevant operators
are neglected (\ie\ operators with scaling dimensions larger than 2).

\item{a)} The Anisotropic Antiferromagnetic ($XXZ$) \heis\ spin chain:

The best known example is the anisotropic Heisenberg
antiferromagnet (the $XXZ$ model) as a function of anisotropy. The
physics of the Thirring/Luttinger model coincides
with the long distance physics of the $XXZ$ model\refto{lutherpeschel}
if umklapp processes can be neglected \refto{haldane,dennijs}. In this
model spin-$\12$ degrees of freedom are arranged on a linear chain with
$N$ sites. The Hamiltonian is
$$
H=J\sum_{j=1}^N \left( \sigma_1(j)\;\sigma_1(j+1)+ \sigma_2(j)
\;\sigma_2(j+1)+\lambda \;\sigma_3(j)\;  \sigma_3(j+1) \right)
\eqno(6.2)
$$
where $\{\sigma_k\}$ ($k=1,2,3$) are the three Pauli matrices, $J$ is
the exchange constant and $\lambda$ measures the anisotropy ($\lambda=1$
is the Heisenberg antiferromagnet). This model is integrable through the
Bethe Ansatz \refto{bethe,yang}. The identification with the
Thirring/Luttinger model was done by Luther and
Peschel\refto{lutherpeschel} (a detailed argument is reviewed in
references[\cite{emery,book}]). The mapping is done in two steps. First,
a
Jordan-\-Wigner transformation is used to map the spin model to a model
of spinless fermions $c(j)$ defined on the same one-\-dimensional
lattice. The Hamiltonian for the equivalent fermion model turns out to
be
$$
H=\frac{J}{2}\sum_{j=1}^N\left(c^{\dagger}(j)\; c(j+1)+{\rm
h.c.}+2\lambda (n(j)-\12)(n(j+1)-\12)\right)\eqno(6.3)
$$
where $n(j)=c^{\dagger}(j)c(j)$ is the fermion occupation number, and
it is related with the z-component of the spin through
$\sigma_3(j)=2n(j)-1$. Hence, the state with zero (one) fermion at $j$
is the same as the state with one spin down (up) at $j$.
The sector with $S_z=0$ is identified with the half-\-filled sector of
the fermion model. The continuum limit of this fermion model in this
sector is easily seen to be equivalent to the Thirring/Luttinger model
once the Fermi field is split into its left and right components.
For small
$\lambda$ (\ie\ close to the $XY$-\-model limit),
the coupling constant for the $g$ Thirring model is seen to be given by
$g^2=\lambda$ (in units such that $v_F=1$). The relation between the two
coupling constants is not universal and, in general, it is
non-\-linear\refto{lutherpeschel}. The umkalpp operators are irrelevant
close to the $XY$ limit ( small $\lambda$) but become marginal at
$\lambda=1$, the isotropic point. In the Thirring/Luttinger
representation these operators become marginal at $g^2=\pi$. For
$g^2>\pi$ these operators become relevant and cause the system to become
massive.

Since the charge density of the Thirring model is proportional to the
average of the fermion occupation numbers on two consecutive sites
(\ie\ the average of the z-\-components of the spins), the wave
function of the Thirring/Luttinger model in the density
representation has to coincide with the wave function of the
Heisenberg chain in the $\sigma_3$ representation. This is the
conventional representation of the Bethe Ansatz wave functions in
which the coordinates represent spin flips. On the other hand, the
Grassmann representation of the Thirring wave functions have coordinates
which represent {\it fermions}, not spin flips. The elementary
excitations of the \heis\ chain are fermions which, in the spin picture,
are solitons. In section IV we found that the wave functions obtained
through these two pictures yield different exponents.
With this identifications, we can now make use of our results from
section IV and conclude that, for large separation of its coordinates,
the wave function of the $XXZ$ model also has a Jastrow form with the
exponent that we found for the Thirring model. This exponent is accurate
only close to the $XY$ limit.
The behavior of the wave function {\it at} the isotropic
\heis\ limit can be determined as a limit $g^2\to \pi$. The exponent of
the Jastrow factor of the wave function, $\nu=(1+g^2/\pi)$, has the
limiting value $\nu=2$. Interestingly enough this is precisely the same
exponent found by Haldane and Shastri in the $1/|x|^2$ spin chain.

\item{b)} Sutherland's model:

Sutherland's model is a system of non-\-relativistic
spinless fermions, at fixed density, interacting \via\ a pair
interaction potential
$V(|x-x'|)=A/|x-x'|^2$, where $A$ is a coupling constant. In second
quantized form the Hamiltonian for the Fermi field $\psi(x)$ is
$$
H=\int dx \; \frac{1}{2M} (\pd_x\psi^{\dagger}(x))\;(\pd_x\psi(x))+
\12 \int dx \int dx' \rho(x) V(|x-x'|) \rho(x')
\eqno(6.4)
$$
where $\rho(x)=\psi^{\dagger}(x) \psi(x)$ is the density operator and
we have set $\hbar=1$ and $M$ is the mass.

By means of a suitable generalization of Bethe's Ansatz,
Sutherland\refto{sutherland}
was able to find the complete spectrum of this model. The wave functions
turned out to have a very simple form.  For a system with periodic
boundary conditions it is natural to think of the particles as being on
a ring of circumference $L$. For the pair potential
$V(|x|)=\frac{g\pi^2}{L^2}\left[\sin(\frac{\pi |x|}{L})\right]^{-2}$,
which has the required periodicity properties, he found that the wave
function for the ground state in the $N$-\-particle sector has the
Jastrow form
$$
\Psi=\prod_{i>j}\left[\psi(x_i-x_j)\right]^{\lambda}\eqno(6.5)
$$
with
$$
\lambda=\12\left[1+(1+2g)^{\12}\right]\eqno(6.6)
$$

We will use the methods of the previous sections to calculate
the asymptotic long distance behavior of the wave function. Up
to a redefinition of the coupling constant, we will be able to
reproduce Sutherland's result, eq~(6.6).

For a general pair potential
$V(|x-x'|)$, Sutherland's model is equivalent (at low energies) to a
Thirring-like model with Hamiltonian
$$
\eqalign{
H=&\int dx \; \left({\bar \psi}(x)i
\gamma_1 \pd_x\psi(x)-\frac{g}{2} ({\bar
\psi}(x)\gamma_{\mu} \psi(x))^2 \right)+\cr\12 &\int dx \int dx'\;
(\psi^{\dagger}(x) \psi(x))
u(|x-x'|)\;(\psi^{\dagger}(x') \psi(x'))\cr}
\eqno(6.7)
$$
where have separated in the Hamiltonian the interactions into forward
scattering backward scattering processes. The forward  scattering
processes are weighed by the low momentum components of the pair
potential (low compared with $2k_F$) $\frac{1}{v_F}\;V(|x|)=u(|x|)$.
The backward scattering processes, which involve momentum transfers near
$2k_F$, are represented by the Thirring term with the coupling constant
$g^2\approx-\frac{1}{2v_F}{\tilde V}(2k_F)$, where $v_F$ is the Fermi
velocity and ${\tilde V}(2k_F)$ is the Fourier transform of the pair
potential at $2k_F$. This separation is accurate for processes in which
the single particle energies differences are much smaller than an energy
cutoff $D$ in which range the dispersion curve is accurately linear.
This restricts the validity of this identification to the weak coupling
regime. For larger values of the coupling constant, it is necessary to
renormalize the parameters of the Thirring model.

With these
caveats we can, in principle, use the wave functions of section IV, and
use these
results in the long distance limit of the Sutherland model. But,
before we do that, we must find out if the density-\-density coupling
could possible change the results. There are several ways to see that
the pair potential cannot affect the long distance behavior. Firstly, by
direct bosonization, it is easy to show that, for potentials $u(|x|)$
which whose Fourier transform ${\tilde u}(q)$ satisfies  $q\;{\tilde
u}(q) \to 0$ ( as $q\to 0$), the effective bosonized action coincides
with the bosonized action of the Thirring model, up to a finite
renormalization of the Thirring coupling constant (\ie\ the
backscattering amplitude). Instead of following that approach, we can
compute the wave function for the generalized theory of eq~(6.7) using
the same methods as in the previous sections. An explicit calculation
in the density representation yields the wave function for the ground
state to be
$$
|\Psi_0[\rho]|^2=\cN \exp\left\{(1+\frac{g^2}{\pi}) \int
\frac{dp}{2\pi}\; \rho(p)\frac{1}{|p|}\left(1-\frac{{\tilde u}(p)}{\pi
(1+\frac{g^2}{\pi})}\right)^{\12}\rho(-p)\right\}
\eqno(6.8)
$$
where $\rho(p)$ is the Fourier transform of the density. This is the
exact wave function for this model (within the approximation of a
linearized dispersion curve). By direct inspection of this result it is
apparent that the Jastrow exponent $\lambda$ is determined by the low
momentum behavior of the potential ${\tilde u}(q)$. Any delta function
contribution to $u(|x|)$ is a constant term in the Fourier transform at
all momenta. Thus, delta function pieces shift (\ie\ renormalize) {\it
both} the Thirring coupling or backscattering amplitude ($\propto {\tilde
V}(2k_F)$) {\it and} the $q \to 0$ behavior of ${\tilde u}(q)$, the
forward scattering amplitude. Thus, such effects are essentially
unobservable since these effects cancel each other out. If the behavior
${\tilde u}(q)\to {\rm const.}$ is excluded, we have to consider a
polynomial (\ie\ analytic) behavior, such as ${\tilde u}(q)\approx A\;
|q|$. This is in fact the behavior for Sutherland's model where the
constant is $A=-\pi g$ (for $V(|x|)=ga/|x|^2$ and $a$ a short distance
cutoff).

Obviously, this analysis fails if the Fourier transform of the
pair potential is singular in the $q \to 0$ limit in the form ${\tilde
u}(q) \approx A\; /q^{\sigma}$ (with $A<0$). For this type of behavior,
the kernel in the integral of eq~(6.8) behaves, at large separations,
like $-{\rm const.}|x-x'|^{\sigma/2}$. Thus, while the wave function
retains its Jastrow form, the physics is completely different: for all
potentials with $\sigma>0$ (\ie\ they decay more {\it slowly} than
$1/|x|^2$ at long distances), the wave function vanishes like
$\exp(-{\rm const.}|x|^{\sigma/2})$ at long distances. This indicates
that, for these potentials, the spectrum of these systems does not
contain particle-\-like excitations. Indeed, the case $\sigma=2$ (\ie\
${\tilde u}=-e^2/q^2$ corresponds to the one-\-dimensional Coulomb
interaction This case is exactly equivalent to electrodynamics in
one space dimension,the Schwinger model. The Schwinger model is known to
 be a {\it confining} theory and it does not contain any states in its
spectrum with fermionic quantum numbers.

Thus, we conclude that, for a large class of pair potentials,
the wave function at long distances not only has the Jastrow form, but
the Jastrow exponent can be calculated directly from the
Thirring/Luttinger model.
In section
IV we got a wave function with the same Jastrow form but with the
exponent $\lambda_{Th}=1+g^2/\pi$ (here $g$ is the coupling constant of
the Thirring model, \ie\ $g\approx -\frac{1}{2v_F}{\tilde V}(2k_F)$).
In the Sutherland model, the Fourier
transform of the pair potential  $V(|x|)=g/|x|^2$ is ${\tilde
V}(|p|)=-(g/v_F)|p|$. Hence, we must identify the Thirring model
$g^2/\pi$ with Sutherland's coupling constant $g$ (for $v_F=1$). This
identification is valid for $g$ small and, in this limit, the
exponent
of the wave function calculated in the Thirring model agrees with
Sutherland's exponent eq~(6.6). However, what matters here is that the
{\it form} of the wave function is the same. The actual value of the
exponent depends on the identification of the coupling constants, which
is not universal. This problem is well known in the context of the
\heis\ spin chain.

\item{c)} The Haldane-Shastri model:

The Haldane-Shastri (HS) model is a non-local modification of the
Heisenberg spin chain with Hamiltonian
$$
H_{HS}=\sum_{j,k=1}^N J(|j-k|) \left(\sigma_1(j)\;\sigma_1(k)+
\sigma_2(j)
\sigma_2(k)+\lambda \;\sigma_3(j)\;  \sigma_3(k)\right)\eqno(6.9)
$$
with $J(|j-k|)=A/|j-k|^2$ (for $j \not= k$). It has the same physics as
the Heisenberg spin chain except that the Bethe Ansatz has a much
simpler structure: the wave functions have a Jastrow form.
But, since the interactions are non-\-local,
the mapping of the $HS$ model to the Thirring/Luttinger model cannot be
done directly, contrary to what we discussed for the Heisenberg
spin chain.

The $HS$ model can be viewed as a strong coupling limit of a
generalization of a Hubbard model. Consider a system of
spin-\-$\12$ fermions on
a one-\-dimensional lattice of length $N$ with creation and annihilation
operators $c^{\dagger}_{\sigma}(j)$ and $c_{\sigma}(j)$ defined at each
site $j$ ($=1,\ldots,N$) and for each spin orientation
$\sigma=\uparrow,\downarrow$. The Hamiltonian
$$
H=\sum_{\sigma=\uparrow,\downarrow}\sum_{j,k=1}^N \left[t(|j-k|)
c^{\dagger}_{\sigma}(j) c_{\sigma}(k) +{\rm h.c.}\right]+\sum_{j=1}^N U
n_{\uparrow}(j)\;n_{\downarrow}(j)\eqno(6.10)
$$
is a Hubbard model with a non-\-local hopping term with amplitude
$t(|j-k|)$. Standard arguments\refto{emery} show that, for a
half-\-filled system and in the strong coupling limit,
($U\gg \max_{j,k} (t(|j-k|))$, this Hubbard model is also equivalent to
a \heis\ antiferromagnet. The exchange interactions are now also
non-\-local
and have an exchange constant \hfil\break
$J(|j-k|)=[t(|j-k|)]^2/U$.
Hence,
the $J/|x|^2$ interaction of the HS model requires that the equivalent
Hubbard model should have a hopping term $t(|x|)\approx \sqrt{JU}/|x|$.
Hence, our results on wave functions also apply to the long distance
behavior of the spin component of the wave functions of the Hubbard
models.

However, all quantum antiferromagnets in all dimensions and for all
lattices, can also
be regarded as the strong coupling limit of a gauge theory\refto{elbio}.
For a one-\-dimensional chain with exchange constant $J(|x_j-x_k|)$ the
correspondence goes as follows: a set of fermion degrees of freedom are
defined on each site of the chain with the corresponding fermion
creation and annihilation operators, $\psi_{a}^{\dagger}(j)$ and
$\psi_{a}(j)$, $a=1,2$. The spin operators are defined to be $S^+(j)=\12
\epsilon_{ab}\psi_a(j) \psi_b(j)$, $S^-(j)=S^+(j)^{\dagger}$ and
$S_3(j)=\psi_a(j)^{\dagger}\psi_a(j)-1$. These operators are invariant
under the {\it local} $SU(2)$ gauge transformation $\psi_a(j)
\to \psi'_a(j)=U_{ab}(j)\psi_b(j)$. In terms of these fermions, the spin
operators have a local symmetry, the quantum antiferromagnets must be
equivalent to an $SU(2)$ gauge theory. Indeed, it is straightforward to
show that they are the strong coupling limit of an $SU(2)$ lattice gauge
theory\refto{marvin, elbio}. The gauge degrees of freedom are $2\times
2$ $SU(2)$ matrices. In the absence of gauge fields, the free fermions
have a global $U(2)$ invariance. The effect of gauge fields is to set
all $SU(2)$ charges and currents to zero on the physical
state\refto{book}. Hence, in the (na{\"\i}ve) continuum limit, this is a
theory of constrained (two-\-component) Dirac fermions in $1+1$
dimensions with the effective Lagrangian
$\cL={\bar \psi} i\slD \psi$, where $D_{\mu}$ is the covariant
derivative in the fundamental (spinor) representation of local $SU(2)$.
This result can be derived by a straightforward application of the
methods used above. The constraint implies that  the physical degrees
of freedom live in the coset $G=U(2)/SU(2)$. (We have assumed that any
non-\-locality is sufficiently weak so that the mapping holds; it
certainly works for the usual nearest neighbor Heisenberg model and it
applies to the HS model as well). More complex coset models are found
for higher spin representations as well as for groups other than
$SU(2)$ (for a review on quantum chains see reference [\cite{affleck}]).

Since all the local $SU(2)$ currents and densities annihilate the
physical states, it is obvious that they cannot be used to label the
states. Thus, the labels of the wave functions have to be the
eigenvalues of the unconstrained currents. The coset $G=U(2)/SU(2)$
contains a global $U(1)$ subgroup whose generator is, precisely,
$\psi^{\dagger}\psi$ which, as we discussed above, is essentially the
same as the spin operator $S_3$ in the Heisenberg model. Hence, the wave
function of the constrained fermion, on the diagonal generators of
the coset $G=U(2)/SU(2)$, has precisely the same information as the
Bethe wave functions for the Heisenberg model. The results of section V
show that these wave functions also have the Jastrow-\-like factorized
form and that the exponent is also equal to $2$ ( see eq~(5.5)), in
agreement with the results of Haldane and Shastri.

\head{VII. Conclusions}

In this paper we have used path-\-integral methods to derive the
ground state wave functions of a number of two-\-dimensional
fermion field theories and related systems in one-\-dimensional many
body physics. We derived the exact wave function for the
Thirring/Luttinger and Coset fermion models and used these results to
derive the universal behavior of the wave functions of the \heis\
models and of the Sutherland model. We found explicit forms for the wave
functions in the density and in the Grassmann representations. For the
field-\-theoretic models, the wave functions can be calculated exactly.
We showed
that these wave functions always have the Jastrow factorized form and
calculated the exponent. The exponents obtained with our methods agree
with the exponents derived
from the Bethe Ansatz for the Sutherland model and the Haldane-\-Shastri
spin chain. We also showed that they also apply to all the
systems which are in the same universality class. Thus, the long
distance form of the ground state wave function for the nearest neighbor
\heis\ antiferromagnet has the same exponent $\nu=2$ as the HS model.
We also verified that this result also applies to the fully $SU(2)$
invariant $U(2)/SU(2)$ constrained (Coset) fermion field theory.
Also, up to a finite renormalization of the coupling
constant, the Thirring/Luttinger form of the wave function agrees with
the wave functions of both the Sutherland model and of the anisotropic
($XXZ$) antiferromagnet.

Our results confirm that the wave functions of these theories also have
a universal content. In previous studies, the universality was always
investigated in the context of the Green functions. Naturally, the
universality of the Green function requires that the wave function
should also have universal component. However, due to the complexity of
the wave functions, their universality is not usually investigated. Our
methods can be extended with only minor effort to the computation of the
wave functions of the excited states. More interesting, however, is the
study of the properties of the wave functions to more complex theories
with Kac-\-Moody level $k>1$. This work is in progress.

\head{Acknowledgments}

We would like to thank Michael Stone and  Ana Lopez for useful
discussions.
This work was supported in part by the National Science Foundation through
grants No.DMR91-22385 at the University of Illinois, by the
NSF-CONICET International
Cooperation Program through the grant No.INT-8902032
University of Illinois, by CONICET (Argentina) and by CIC (Argentina).

\endpage

\references

\refis{rajaraman} R.Rajaraman, {\it ``Solitons and Instantons"},
North-Holland, Amsterdam (1987).

\refis{leshouches} For a review on the Bethe Ansatz see {\it Recent
Advances in Field Theory and Statistical Mechanics}, Les Houches Summer
School 1982, Session XXXIX, J-B.~Zuber and R.~Stora Editors, North
Holland (Amsterdam)(1984).

\refis{sutherland} B.~Sutherland, \jmp 12, 246, 1971; \pra 4, 2019,
1971.

\refis{shastri} B.~S.~Shastri, \prl 60, 639, 1988.

\refis{haldaner^2} F.~D.~M.~Haldane, \prl 60, 635, 1988.

\refis{jackiw} See, for instance, R.~Jackiw , ``Schr\"odinger Picture
for Boson and Fermion Quantum Field Theories", in ``{\it Mathematical
Quantum Field Theory and Related Topics}", page 107, Proceedings of the
1987
Montr\'eal Conference of the Canadian Mathematical Society, Conference
Proceedings Vol.~9, J.~Feldman and L.~Rosen Editors, American and
Canadian Mathematical Societies (1987).

\refis{wheeler} J.~A.~Wheeler, ``{\it Geometrodynamics}", Academic Press
(New York, 1962).

\refis{symanzik} K.~ Symanzik, \np {\bf B}190, 1, 1981.

\refis{luscher} M.~Luscher, \np {\bf B}254, 52, 1985.

\refis{landauhelium} L.~D.~Landau, \journal Jour.~Physics, 5, 71, 1941.

\refis{dashen} R.~F.~Dashen and D.~H.~Sharp, \pr 165, 1857, 1968.

\refis{haldane} F.~D.~M.~Haldane, \prb 25, 4925, 1982.

\refis{lutherpeschel} A.~Luther and I.~Peschel, \prb 12, 3908, 1975.

\refis{emery} V.~J.~Emery in {\it Highly Conducting One Dimensional
Solids}, J.~T.~Devreese, R.~P.~Evrard and V.~E.~van Doren editors,
Plenum Press (New York)(1979).

\refis{dennijs} M.~P.~M.~ den Nijs, \prb 23, 6111, 1981.

\refis{yang} C.~N.~Yang and C.~P.~Yang, \jmp 10, 1115, 1969.

\refis{bethe} H.~Bethe, \journal Z.~Physik, 71, 205, 1931.

\refis{elbio} E.~Dagotto, E.~Fradkin and A.~Moreo, \prb 38, 2926, 1988.

\refis{marvin} S.~Drell, M.~Weinstein and S.~Yankielowicz, \prd 14,
1627, 1976.

\refis{book} E.Fradkin, {\it ``Field Theories of Condensed Matter
Systems"}, Addison-Wesley, Redwood City (1991).

\refis{anawave} Ana Lopez and Eduardo Fradkin, `` Universal Properties
of the Wave Functions of Fractional Quantum Hall Systems ",  Urbana
Preprint, March 1992.

\refis{wavefun} E.~Fradkin, ``Wave Functionals for Field Theories and
Path Integrals", Urbana Preprint, May 1992.

\refis{affleck} I.~K.~Affleck, \np {\bf B}265, 409, 1986.

\refis{enri} D.~Cabra, E.~Moreno and M.~C.~von Reichenbach, \journal Int. Jour.
of Mod. Phys. A , 5, 2313, 1990; E.~Fradkin, C.~M.~Naon and F.~A.~Schaposnik,
\pl 200B, 95, 1987.

\refis{goddard} P.~Goddard, A.~Kent and D.~Olive, \pl 15B, 88, 1985; \cmp 103,
105, 1985.

\refis{polyakov} A. ~M.~Polyakov, Lectures
at Les Houches Session XLIX 1988, eds. E. Brezin and J. Zinn Justin, Elsevier,
1989.

\refis{polyakov2} A.~M.~Polyakov and P.~B.~Wiegman, \pl 131B, 121, 1983; \pl
141B, 223, 1984.

\refis{fidel} See for example F.~A.~Schaposnik, Lectures at the IV Swieca
School, Particles and Fields, 1987; eds. A.da Silva et al, World.Sci.

\endreferences

\endit